\documentclass[11pt]{article}

\usepackage{graphicx}
\usepackage{amsmath, amssymb}

\newcommand{\vct}[1]{\boldsymbol{#1}}
\newcommand{\ket}[1]{|\,#1\,\rangle}
\newcommand{\mtrix}[3]{\langle \,#1\,|\,#2\,|\,#3\,\rangle}
\newcommand{\pf}{p_{\mathrm{F}}}
\newcommand{\epf}{E_{p_{\mathrm{F}}}}
\newcommand{\ef}{E_{\mathrm{F}}}
\newcommand{\vf}{v_{\mathrm{F}}}
\newcommand{\nf}{N_{\mathrm{F}}}
\newcommand{\calV}{\mathcal{V}}
\newcommand{\rph}{R^{\mathrm{ph}}}
\newcommand{\rhn}{R^{\mathrm{h\bar{n}}}}
\newcommand{\rpn}{R^{\mathrm{p\bar{n}}}}
\newcommand{\rnn}{R^{\mathrm{n\bar{n}}}}
\newcommand{\nbar}{\mathrm{\overline{N}}}
\newcommand{\srpa}{S_{\mathrm{RPA}}}
\newcommand{\sha}{S_{\mathrm{H}}}
\newcommand{\piH}{\varPi_{\mathrm{H}}}
\newcommand{\grH}{G_{\mathrm{H}}}
\newcommand{\rNSA}{\mathrm{NSA}}
\newcommand{\rNFA}{\mathrm{NFA}}
\newcommand{\rFull}{\mathrm{Full}}
\newcommand{\dd}{\mathrm{d}}
\newcommand{\hf}{\!\ }

\begin{document}

\baselineskip=14pt

{
\noindent
\LARGE\textbf{%
Roles of Antinucleon Degrees of Freedom in the\\[4pt]
Relativistic Random Phase Approximation}
}

\vspace{\baselineskip}

\noindent
Haruki Kurasawa$^{1}$ and Toshio Suzuki$^{2}$

\vspace{0.5\baselineskip}

\noindent
$^{1}$\ \parbox[t]{\textwidth}{Department of Physics, Graduate School
of Science, Chiba University, Chiba\\[2pt] 263-8522, Japan
}

\vspace{0.5\baselineskip}

\noindent
$^{2}$\ Emeritus Professor, University of Fukui, Fukui 910-8507, Japan

\begin{center}
 
\parbox[t]{13cm}{
\small

\baselineskip=12pt

\noindent
Roles of antinucleon degrees of freedom in the relativistic random
phase approximation(RPA) are investigated. The energy-weighted sum
of the RPA transition strengths is expressed in terms of the double
commutator between the excitation operator and the Hamiltonian,
as in nonrelativistic models. The commutator, however, should not
be calculated with a usual way in the local field theory, because,
otherwise, the sum vanishes. The sum value obtained correctly from
the commutator is infinite, owing to the Dirac sea. Most of the
previous calculations takes into account only a part of the 
nucleon-antinucleon states, in order to avoid the divergence problems. 
As a result, RPA states with negative excitation energy appear,
which make the sum value vanish. Moreover, disregarding the divergence
changes the sign of nuclear interactions in the RPA equation which
describes the coupling of the nucleon particle-hole states with the 
nucleon-antinucleon states. Indeed, excitation energies of the
spurious state and giant monopole states in the no-sea approximation
are dominated by those unphysical changes. The baryon current
conservation can be described without touching the divergence
problems. A schematic model with separable interactions is presented,
which makes the structure of the relativistic RPA transparent.
}
\end{center}

\vspace{0.5\baselineskip}

\section{Introduction}

It has been shown by many authors\cite{sw,pr,lnnm,cpc}
that relativistic models, 
assuming nuclei to be composed of Dirac particles and various mesons,
work well phenomenologically to reproduce nuclear static and dynamic properties.
In most of calculations, however, antinucleon($\nbar$) degrees of freedom
are not fully taken into account, in spite of the fact that they are one of
the main differences between relativistic and non-relativistic models.
The reason why the full space is not used is because there exist 
divergence problems\cite{sw,chin,ks1} which are not yet handled
with a proper way for finite nuclei.

Although all the space was not included in the previous calculations,
a part of $\nbar$ degrees of freedom was taken into account,
aiming to keep some fundamental principles and to reproduce gross
properties of nuclei.

For example, in the random phase approximation (RPA),
the baryon current conservation required some excitations
of antinucleons\cite{chin,ks1,ks2}.
It was necessary for description of the center of mass motion
to have the Landau-Migdal parameter $F_1$ taking into account
a part of N(nucleon)-$\nbar$ states\cite{ks2,nks}.
In the same way, the spurious states in RPA\cite{srm,df} 
was described with those N-$\nbar$ excitations 
in addition to nucleon particle-hole states.
Furthermore, an abnormal enhancement of the isoscalar magnetic moment
in the Hartree approximation demanded corrections
from the N-$\nbar$ excitations through $F_1$ \cite{ks4}.

For the above reasons, the two ways to include the $\nbar$ space,
avoiding the divergence problem, were proposed.
The one is the no free term approximation(NFA) which simply neglects 
the divergent part of the RPA response functions.
The remaining part is composed of transitions of antinucleons
to the Fermi sea which are in fact Pauli-blocked\cite{chin,ks1}.
The other is the no-sea approximation(NSA) to assume 
that all the $\nbar$ states are empty.
There is no divergence problem in this case also, 
but transitions of nucleons in the Fermi sea to the $\nbar$ states
are permitted with negative excitation energies\cite{df}.
Even though both methods look unreasonable,
they have been widely used, since experimental
data are well reproduced phenomenologically\cite{lnnm, piek,mgwvr,mcg}.
In fact, as will be seen later explicitly,
NFA and NSA are equivalent to each other in RPA\cite{df}. 

The purpose of the present paper is to investigate the structure of
the relativistic RPA in detail, and in particular,
a role of $\nbar$ degrees of freedom there.
The relativistic RPA can be developed almost in the same way
as non-relativistic ones. It will be shown, however, that 
a careful treatment is required for $\nbar$ 
degrees of freedom.
They cause the divergence problems, but cannot be simply ignored
as in NFA and NSA.

It will be shown that the energy-weighted sum of the RPA transition strengths
is expressed formally in terms of the double commutator between
the transition operator and the Hamiltonian.
This RPA sum rule is the same as in nonrelativistic case. 
The commutator, however, should not be calculated employing
a usual rule in the local field theory. Otherwise, the sum vanishes, 
in spite of the fact that it should be positive definite.
The correct calculation of the commutator
gives the sum value to be infinite, because of N-$\nbar$ 
excitations.

In NFA and NSA,   
RPA states with negative excitation-energy appear,
in addition to those with positive energy. This is due to
the fact that the full N-$\nbar$ excitations
are not take into account. 
As a result, the energy-weighted sum value 
of the excitation strengths vanishes.
Moreover, disregard of the divergence in NSA and NFA 
gives rise to other unphysical results.
The sign of the nuclear interactions 
is changed in RPA equation relevant to the remaining
N-$\nbar$ states.
Because of this fact,
attractive (repulsive)
forces work as repulsive (attractive) ones
in the coupling of nucleon particle-hole states
with N-$\nbar$ states considered in NFA and NSA.
These effects are not negligible,
but dominate the excitation energies of some low lying states.
For example, the previous numerical calculations could reproduce
a spurious state\cite{srm,df} and giant monopole states\cite{piek,mgwvr,mcg},
invoking those unphysical effects hidden in the RPA equation.

The above defects of NFA and NSA have not been investigated
explicitly so far,
since the structure of the relativistic RPA formulae may be
rather complicated, compared with non-relativistic ones.
In the next section, we will briefly review the relativistic RPA
focusing on roles of N-$\nbar$ states in various approximations.  
In \S\ref{ews}, the energy-weighted sum of the RPA excitation strengths
is discussed, according to recent new insight for the long-standing
problem on the relativistic sum rule\cite{ks5}.
In \S\ref{ce} and \S\ref{ss}, we will explore in detail  
why NFA and NSA seemed to describe well 
the continuity equation and the spurious state without the full
N-$\nbar$ excitations.

It is well-known that a schematic model with separable interactions  
helps us to understand the structure of nonrelativistic RPA\cite{rowe}.
In \S\ref{sm}, a similar model will be introduced for the relativistic RPA,
in order to make clear the present discussions.
The final section is devoted to a brief
summary of the present paper.

\section{Relativistic RPA}\label{r}

The relativistic RPA was formulated
in various ways\cite{sw,chin,ks1,srm,df}.
Taking notice of the dependence on N-$\nbar$ states,
let us briefly review it.  

We begin with the Hartree polarization function for arbitrary
$4\times 4$ matrices, $A$ and $B$\ \cite{ks1},
\begin{equation}
i\piH(A(1),B(2))
= 
\mtrix{\hf}{\mathrm{T}\Bigl(\overline{\psi}(1)A\psi(1)\,
\overline{\psi}(2)B\psi(2)\Bigr)}{\hf},\label{hpf1}
\end{equation}
defined for the Hartree ground state $\ket{\hf}$.
The baryon field $\psi(x)$ is written using the complete set of
the eigenfunction, $\varphi_\alpha$, of Hartree Hamiltonian, 
\begin{equation}
 \psi(x)
=\sum_{\alpha}\varphi_\alpha(\vct{x})e^{-iE_\alpha t}a_\alpha,
\end{equation}
where $E_\alpha$  denotes the eigenvalue of $\varphi_\alpha$,
and $a_\alpha$ plays a role of
the annihilation or creation operator of N and
$\nbar$, satisfying
$\{\,a_\alpha\,,\,a_\beta^\dagger\,\} =\delta_{\alpha\beta}$, etc..
Then, together with the closure property of $\varphi_\alpha$,
\begin{equation}
 \sum_{\alpha}
\varphi_\alpha(\vct{x})
\varphi^\dagger_\alpha(\vct{x}')=\delta(\vct{x}-\vct{x}'),\label{comp}
\end{equation}
the baryon field satisfies the anti-commutation relation,
\begin{equation}
\{\,\psi_a(x_0,\vct{x})\,,\,\psi_b^\dagger(x_0,\vct{x}')\,\}
 =\delta_{ab}\,\delta(\vct{x}-\vct{x}'),\label{anti_com}
\end{equation}
where $a$ and $b$ denote the Dirac matrix indices.
Since the simple interactions have rather advantage for our purpose
to investigate the detail of the relativistic RPA structure, 
we assume, throughout the present paper,
the $\sigma$-$\omega$ model\cite{sw},
which provides us with the Hartree Hamiltonian as
\begin{equation}
h(x)=-\,i\vct{\alpha}\!\cdot\!\vct{\nabla}+\gamma_0M
 +V_{\mathrm{H}}(x)\,,\quad
V_{\mathrm{H}}(x)=\gamma_0\varSigma(x).\label{hpot}
\end{equation}
In the above equation,
we have employed the following abbreviation for the potential,
\begin{equation}
\varSigma(x)
=
V_\sigma(x)+\gamma^\mu V_\mu(x)
\end{equation}
with
\begin{align}
 V_\sigma(x)&=
 g_\sigma^2
 \int\!\dd^4y\,D_{\sigma}(x-y)
 \,\mtrix{\hf}{\overline{\psi}(y)\psi(y)}{\hf}, \label{def_v_sigma}\\
 V_\mu(x)&=
 g_\omega^2
 \int\!\dd^4y\,D_{\omega}(x-y)
 \,\mtrix{\hf}{\overline{\psi}(y)\gamma_\mu\psi(y)}{\hf}, 
 \label{def_v_omega}
\end{align}
$g_\sigma$ and $g_\omega$ being the coupling constants,
and $D_\sigma$ and $D_\omega$ denoting the $\sigma$- and 
$\omega$-meson propagators, respectively\cite{ks1},
\begin{align}
 D_\sigma(x)
&= 
 \int\!\frac{\dd k_0}{2\pi}
 e^{-ik_0x_0}D_\sigma(k_0,\vct{x}) ,\quad
 D_\sigma(k_0,\vct{x})=
 \int\!\frac{\dd^3k}{(2\pi)^3}
 \,\frac{e^{i\vct{k}\cdot\vct{x}}}{k_\mu k^\mu-m_\sigma^2+i\varepsilon},
 \label{d_sigma2} \\
 D_\omega(x)
&=
 \int\!\frac{\dd k_0}{2\pi}
 e^{-ik_0x_0}D_\omega(k_0,\vct{x}) , \quad
 D_\omega(k_0,\vct{x})=
 -\int\!\frac{\dd^3k}{(2\pi)^3}
 \,\frac{e^{i\vct{k}\cdot\vct{x}}}{k_\mu k^\mu-m_\omega^2+i\varepsilon}.
 \label{d_omega2}
\end{align}
Here, $m_\sigma$ and $m_\omega$ represent the masses
of the $\sigma$- and $\omega$-mesons.

The calculation of Eq.(\ref{hpf1}) gives\cite{ks1},
\begin{equation}
  \piH(A(1),B(2))
=
\frac{1}{2\pi}\int\!\dd\omega\,
e^{-i\omega(t_1-t_2)}
\piH(A(\vct{x}_1),B(\vct{x}_2),\omega),\label{hpf}
\end{equation}
where we have defined
\begin{equation}
 \piH(A(\vct{x}_1),B(\vct{x}_2),\omega)
=
\sum_{\alpha\beta}
\overline{\varphi}_\beta(\vct{x}_1)A\varphi_\alpha(\vct{x}_1)\,
\overline{\varphi}_\alpha(\vct{x}_2) B \varphi_\beta(\vct{x}_2)
\,(\piH)_{\alpha\beta}(\omega) 
\label{pi_h_x1_x2} 
\end{equation}
with
\begin{equation}
 (\piH)_{\alpha\beta}(\omega)
=
\frac{N_{\alpha\beta}}{\omega-E_{\alpha\beta}+i\varepsilon}
-\frac{N_{\beta\alpha}}{\omega-E_{\alpha\beta}-i\varepsilon} 
\,,\qquad
N_{\alpha\beta}=\left(1-\theta_\alpha\right)\theta_\beta.\label{nalphabta}
\end{equation}
In this equation,
$E_{\alpha\beta}$ stands for the excitation energy, $E_\alpha-E_\beta$,
and $N_{\alpha\beta}$ implies the transition from the occupied state to
unoccupied state as,
\begin{equation}
\theta_\alpha=
\left\{
\begin{array}{ll}
1\,, & \alpha\,:\,\mbox{occupied state in }\ket{\hf}, \\[2pt]
0\,, & \alpha\,:\,\mbox{unoccupied state in }\ket{\hf} .
\end{array}
\right. \label{def_theta_alpha}
\end{equation}
Defining the Fermi energy by $\ef$,
it is convenient to employ the following notations for $N_{\alpha\beta}$,
\begin{equation}
 p_\alpha=\theta(E_\alpha-\ef)\,,\qquad 
 h_\alpha=\theta(E_\alpha)\,\theta(\ef-E_\alpha)\,,\qquad
 \bar{n}_\alpha=\theta(-E_\alpha),\label{phnbar}
\end{equation}
where $p_\alpha$, $h_\alpha$, and $\bar{n}_\alpha$ indicate the particle,
hole and $\nbar$ states, respectively.
In the full space calculation, the occupied states are expressed
by ($h+\bar{n}$), and unoccupied states by $p$,
so that $N_{\alpha\beta}$ is written as
$N_{\alpha\beta} = p_\alpha(h_\beta+\bar{n}_\beta)$.
In NSA where the $\nbar$ states are empty, we have
$N_{\alpha\beta} = (p_\alpha+\bar{n}_\alpha)h_\beta$.
In writing
$N_{\alpha\beta}
=p_{\alpha}h_\beta-h_\alpha\bar{n}_\beta+(p_\alpha+h_\alpha)\bar{n}_\beta$ 
for the full space,
NFA neglects the last term which
describes the vacuum polarization to produce the divergence\cite{chin,ks1}. 
In the approximation neglecting all the $\nbar$
states(No$\nbar$),
$N_{\alpha\beta}$ is simply given by $p_\alpha h_\beta$.
Thus, the difference between various approximations can be represented 
by $N_{\alpha\beta}$, reading it as
\begin{equation}
N_{\alpha\beta}
=\left\{
\begin{array}{ll}
p_\alpha h_\beta+p_\alpha\bar{n}_\beta\, & (\mbox{Full}), \\[2pt]
p_\alpha h_\beta+\bar{n}_\alpha h_\beta\, & \mbox{(NSA)}, \\[2pt]
p_\alpha h_\beta-h_\alpha\bar{n}_\beta\, & \mbox{(NFA)}, \\[2pt]
p_\alpha h_\beta\, & (\mathrm{No}\nbar)\,.
\end{array}
\right.
\label{pi0_3}
\end{equation}

We define the inverse of $\piH$ by $\piH^{-1}$ as
\[
 \sum_{\alpha''\beta''}
(\piH^{-1})_{\alpha\beta,\alpha''\beta''}
(\piH)_{\alpha''\beta'',\alpha'\beta'}
=I_{\alpha\beta,\alpha'\beta'},
\]
where the following abbreviations are used
\[
(\piH)_{\alpha\beta,\alpha'\beta'}
=\delta_{\alpha\alpha'}\delta_{\beta\beta'}
(\piH)_{\alpha\beta}\,,\qquad
 I_{\alpha\beta,\alpha'\beta'}=
\delta_{\alpha\alpha'}\delta_{\beta\beta'}
\Bigl(
N_{\alpha\beta}^2+N_{\beta\alpha}^2
\Bigr).
\]
For $N_{\alpha\beta}^2+N_{\beta\alpha}^2=1$, we obtain
\begin{equation}
  (\piH^{-1}(\omega))_{\alpha\beta,\alpha'\beta'}
=
\delta_{\alpha\alpha'}\delta_{\beta\beta'}
\left( N_{\alpha\beta}-N_{\beta\alpha} \right)
\left(\omega-E_{\alpha\beta}\right).
\label{pi0_inv}
\end{equation}
Because of $p_\alpha+h_\alpha+\bar{n}_\alpha=1$,
the above ($N_{\alpha\beta}-N_{\beta\alpha}$) 
is given for each approximation as
\begin{equation}
  N_{\alpha\beta}-N_{\beta\alpha}
=\left\{
\begin{array}{ll}
(h_\beta+\bar{n}_\beta)-(h_\alpha+\bar{n}_\alpha)\, & \mbox{(Full),}  \\[2pt]
h_\beta-h_\alpha\, & \mbox{(NSA, NFA),} \\[2pt]
p_\alpha h_\beta-p_\beta h_\alpha \, & (\mathrm{No}\nbar).\,
\end{array}
\right.
\label{nab-nba}
\end{equation}
Unlike $\piH$, the inverse $\piH^{-1}$ of NSA is the
same as for NFA.
This fact makes NSA and NFA are equivalent
to each other in RPA, as seen later.

The RPA polarization function is written
in terms of $\piH$\cite{ks1},
\begin{align}
 \varPi(A(1),B(2))&=
 \piH(A(1),B(2))\nonumber\\
 &\phantom{=} 
+ g_\eta^2
 \int\!\dd^4x\,\dd^4y\,
 \piH(A(1),\varGamma^\eta(x))
 D_{\eta}(x-y)
 \varPi(\varGamma_{\eta}(y),B(2)),
 \label{def_rpa_pola} 
\end{align}
where $\eta=-1$ for $\sigma$
and, hence, $\varGamma_{\eta}=1$ for $\eta=-1$, and $\gamma_\mu$
for $\eta=\mu$.
As in $\piH$, we  write Eq.(\ref{def_rpa_pola})  in the form:
\begin{equation}
\varPi(A(1),B(2))
=\frac{1}{2\pi}\int\!\dd\omega\,
e^{-i\omega(t_1-t_2)}
\varPi(A(\vct{x}_1),B(\vct{x}_2),\omega),\label{rpa_pi_x1_x2}
\end{equation}
with
\begin{align*}
\varPi(A(\vct{x}_1),B(\vct{x}_2),\omega) 
 &=
 \piH(A(\vct{x}_1),B(\vct{x}_2),\omega) \\
 & +
 g_\eta^2\!
 \int\!\dd^3x\, \dd^3y\,
 \piH(A(\vct{x}_1),\varGamma^\eta(\vct{x}),\omega)
 D_{\eta}(\omega,\vct{x}-\vct{y})
 \varPi(\varGamma_{\eta}(\vct{y}),B(\vct{x}_2),\omega). 
\end{align*}
The above $\varPi(A(\vct{x}_1),B(\vct{x}_2),\omega)$  is described as
\begin{equation}
 \varPi(A(\vct{x}_1),B(\vct{x}_2),\omega)
=
\sum_{\alpha\beta\alpha'\beta'}
\overline{\varphi}_\beta(\vct{x}_1)A\varphi_\alpha(\vct{x}_1)\,
\varPi_{\alpha\beta,\alpha'\beta'}(\omega)
\,\overline{\varphi}_{\alpha'}(\vct{x}_2)B\varphi_{\beta'}(\vct{x}_2)
\label{def_pi_omega},
\end{equation}
where we have defined
\begin{equation}
\varPi_{\alpha\beta,\alpha'\beta'}(\omega)
=
\delta_{\alpha\alpha'}\delta_{\beta\beta'}
(\piH)_{\alpha\beta}(\omega)
+\sum_{\alpha''\beta''} (\piH)_{\alpha\beta}(\omega)
V_{\alpha\beta'',\beta\alpha''}(\omega)
\varPi_{\alpha''\beta'',\alpha'\beta'}(\omega),
\label{pl_eq1}
\end{equation}
using the notation
\[
 V_{\alpha\beta',\beta\alpha'}(\omega) 
=g_\eta^2\int\!\dd^3x_1\,\dd^3x_2\,
\overline{\varphi}_\alpha(\vct{x}_1)\varGamma^\eta\varphi_\beta(\vct{x}_1)
D_{\eta}(\omega,\vct{x}_1-\vct{x}_2)
\,\overline{\varphi}_{\beta'}(\vct{x}_2)
\varGamma_{\eta}\varphi_{\alpha'}(\vct{x}_2).
\]
This is also written as
\begin{equation}
V_{\alpha\beta',\beta\alpha'}(\omega)
=
\mtrix{\alpha,\beta'}{\,\calV_{12}(\omega)\,}{\beta,\alpha'},
\label{def_two_body_int}
\end{equation}
with
\[
\calV_{12}(\omega)=
g_\eta^2(\gamma_0\varGamma^\eta)_1
(\gamma_0\varGamma_{\eta})_2
D_{\eta}(\omega,\vct{x}_1-\vct{x}_2).
\]
In relativistic models, nuclear interactions contain $\omega$-dependence
coming from Eqs.(\ref{d_sigma2}) and (\ref{d_omega2}).
For later discussions of the relativistic RPA, however, we have to neglect
this retardation effect, as in nonrelativistic RPA.
Fortunately, their contributions to 
the interactions have been shown to be negligible in Ref.\cite{ks6},
when $m_\sigma, m_\omega > \omega$.
Hence, from now on we will develop the RPA, assuming $\calV_{12}(0)$.

Eq.(\ref{pl_eq1})  is formally described as follows,  
\begin{equation}
 \varPi(\omega)
=
\piH(\omega)
+\piH(\omega) V\varPi(\omega)\,,\qquad
(\piH)_{\alpha\beta,\alpha'\beta'}(\omega)
\equiv \delta_{\alpha\alpha'}\delta_{\beta\beta'}
(\piH)_{\alpha\beta}(\omega).
\label{pl_eq2}
\end{equation}
Then, the inverse of $\varPi(\omega)$ can be written as
$
\varPi^{-1}=\piH^{-1}-V 
$ 
with
\begin{equation}
 (\varPi^{-1}(\omega))_{\alpha\beta,\alpha'\beta'}
=
\delta_{\alpha\alpha'}\delta_{\beta\beta'}
\left( N_{\alpha\beta}-N_{\beta\alpha} \right)
\left(\omega-E_{\alpha\beta}\right)
-V_{\alpha\beta',\beta\alpha'}.\label{inverse_rpa}
\end{equation}

Now let us define the eigenvector  $C^{(n)}_{\alpha\beta}$
of the above equation\cite{fw},
\[
 \varPi^{-1}(\omega_n)\,C^{(n)}=0,
\]
which is
\begin{equation}
\left( N_{\alpha\beta}-N_{\beta\alpha} \right)E_{\alpha\beta}
C^{(n)}_{\alpha\beta}
+
\sum_{\alpha'\beta'}V_{\alpha\beta',\beta\alpha'}
C^{(n)}_{\alpha'\beta'}=
\left( N_{\alpha\beta}-N_{\beta\alpha} \right)
\omega_n C^{(n)}_{\alpha\beta}.
\label{eq_coef_c}
\end{equation}
In writing 
\begin{equation}
C^{(n)}_{\alpha\beta}=N_{\alpha\beta}X^{(n)}_{\alpha\beta}
+N_{\beta\alpha}Y^{(n)}_{\beta\alpha},\label{eigenvector}
\end{equation}
the coupled equations are obtained,
\begin{equation}
\begin{split}
E_{\alpha\beta} X^{(n)}_{\alpha\beta}
+
\sum_{\alpha'\beta'}N_{\alpha'\beta'}
\Bigl(V_{\alpha\beta',\beta\alpha'}
X^{(n)}_{\alpha'\beta'}
+
V_{\alpha\alpha',\beta\beta'}
Y^{(n)}_{\alpha'\beta'}\Bigr)
&= \omega_n X^{(n)}_{\alpha\beta}, \\
E_{\alpha\beta} Y^{(n)}_{\alpha\beta}
+
\sum_{\alpha'\beta'} N_{\alpha'\beta'}
\Bigl(V_{\beta\alpha',\alpha\beta'}
Y^{(n)}_{\alpha'\beta'}
+
V_{\beta\beta',\alpha\alpha'}
X^{(n)}_{\alpha'\beta'}\Bigr)
&=-\,\omega_n Y^{(n)}_{\alpha\beta} .
\end{split}
\label{rel_rpa_eq0}
\end{equation}
Employing the abbreviations 
\begin{equation}
\begin{split}
  A_{\alpha\beta,\alpha'\beta'}
&=
\delta_{\alpha\alpha'}\delta_{\beta\beta'}N_{\alpha\beta}E_{\alpha\beta}
+N_{\alpha\beta}V_{\alpha\beta',\beta\alpha'}N_{\alpha'\beta'}, \\ 
B_{\alpha\beta,\alpha'\beta'}
&=
N_{\alpha\beta}V_{\alpha\alpha',\beta\beta'}N_{\alpha'\beta'},
\end{split}
\label{def_rrpa_ab}
\end{equation}
Eq.(\ref{rel_rpa_eq0}) provides us with the relativistic RPA equation
of the form:
 \begin{equation}
\sum_{\alpha'\beta'}
\left(
\begin{array}{cc}
A_{\alpha\beta,\alpha'\beta'} &
B_{\alpha\beta,\alpha'\beta'} \\[4pt]
B_{\alpha\beta,\alpha'\beta'}^\ast &
A_{\alpha\beta,\alpha'\beta'}^\ast
\end{array}
\right) 
\left(
\begin{array}{c}
X_{\alpha'\beta'}^{(n)} \\[6pt]
Y_{\alpha'\beta'}^{(n)}
\end{array}
\right)
=\omega_n\,N_{\alpha\beta}
\left(
\begin{array}{c}
X_{\alpha\beta}^{(n)} \\[6pt]
-\,Y_{\alpha\beta}^{(n)}
\end{array}
\right). \label{rel_rpa_eq}
\end{equation}
When the $\nbar$ degrees of freedom in $N_{\alpha\beta}$
 are neglected,
the above equation reduces to the well-known non-relativistic RPA
equation\cite{rowe,th}.

The relationship between the Full and NFA can be seen 
in Eq.(\ref{eigenvector}).
Comparing $N_{\alpha\beta}$ of NFA with the one of the Full case, 
the part of the N-$\nbar$ excitations in NFA has a minus sign, 
as $-h_\alpha\bar{n}_\beta$, because of neglecting the divergent part
as mentioned before.
This additional sign induces unphysical effects in NFA 
that attractive(repulsive) forces work as repulsive(attractive) 
ones in the $h_\alpha\bar{n}_\beta$-dependent part of the RPA equation, 
Eq.(\ref{rel_rpa_eq0}).

It is also seen from Eq.(\ref{eigenvector}) that NSA is equivalent to NFA.  
In writing Eq.(\ref{eigenvector}) explicitly for NSA and NFA as 
\begin{align*}
 C^{\rNSA}_{\alpha\beta}
 &= 
 \bigl(p_\alpha h_\beta+\bar{n}_\alpha h_\beta \bigr)X^{\rNSA}_{\alpha\beta} 
 +
 \bigl(p_\beta h_\alpha +\bar{n}_\beta h_\alpha \bigr)Y^{\rNSA}_{\beta\alpha}, 
 \\[4pt]
 C^{\rNFA}_{\alpha\beta}
 &= 
 \bigl(p_\alpha h_\beta-h_\alpha \bar{n}_\beta \bigr)X^{\rNFA}_{\alpha\beta} 
 +
 \bigl(p_\beta h_\alpha-h_\beta \bar{n}_\alpha \bigr)Y^{\rNFA}_{\beta\alpha},
\end{align*}
the replacements
\[
 X^{\rNSA}_{ph}=X^{\rNFA}_{ph}\,,\quad
 X^{\rNSA}_{\bar{n}h}=-\,Y^{\rNFA}_{h\bar{n}}\,,\quad
 Y^{\rNSA}_{ph}=Y^{\rNFA}_{ph}\,,\quad
 Y^{\rNSA}_{\bar{n}h}=-\,X^{\rNFA}_{h\bar{n}}
\]
lead to
\[
 C^{\rNSA}_{\alpha\beta}
=C^{\rNFA}_{\alpha\beta}.
\]
Hence, NSA yields same unphysical change of the sign in the interactions,
and the same eigenvalues, as in NFA.

We add a few comments which we need for later discussions.
First, the complex conjugate of Eq.(\ref{eq_coef_c}) implies that
$C^{(n)\,\ast}_{\beta\alpha}$ is also its solution with the 
eigenvalue to be $-\,\omega_n$.
Second,
the orthogonality and normalization of the eigenvectors 
are written as
\begin{equation}
 \sum_{\alpha\beta}
 \left(  N_{\alpha\beta}-N_{\beta\alpha} \right)
C^{(n)\ast}_{\alpha\beta}\,C^{(n')}_{\alpha\beta}
=\lambda_n\delta_{nn'}\ (\lambda_n =\pm 1).
\label{oth_coef_c1}
\end{equation}
The eigenvectors $C^{(n)}_{\alpha\beta}$ with  $\lambda_n=1$ 
describe the RPA excited states. 
Thus, the norm of the eigenvectors in NSA is also the same as in NFA. 
Third, the closure relation is given by
\begin{equation}
 \sum_n
\lambda_n
C_{\alpha\beta}^{(n)} C^{(n)\ast}_{\alpha'\beta'}
=\delta_{\alpha\alpha'}\delta_{\beta\beta'}
\left( N_{\alpha\beta}-N_{\beta\alpha} \right).
\label{comp_cc}
\end{equation}
Finally, we note that Eqs.(\ref{rel_rpa_eq}) and (\ref{oth_coef_c1})
provide us with
\begin{equation}
 \omega_n\lambda_n=
\left(\,X^{(n)\dagger}\ 
Y^{(n)\dagger} \right)
{\cal M}
\left(
\begin{array}{c}
X^{(n)} \\[6pt]
Y^{(n)}
\end{array}
\right)\,,\qquad
{\cal M}
=\left(
\begin{array}{cc}
A &
B \\[4pt]
B^\ast &
A^\ast
\end{array}
\right). 
\label{def_matrix_m}
\end{equation}
In non-relativistic models, we usually obtain $\omega_n\geq 0$
for $\lambda_n=1$\cite{rowe,th}.
In NSA and NFA, however, it is not necessary for ${\cal M}$ to be
positive definite, 
so that there may be negative eigenvalues $\omega_n<0$ even for $\lambda_n=1$.

\section{Energy-weighted sum of the excitation strengths}\label{ews}

One of the reason why RPA has extensively been used in non-relativistic
nuclear models is because the sum of the energy-weighted 
strengths for exciting the RPA states is constrained in a simple way
by the employed Hartree or Hartree-Fock basis.
Once we have the double commutator of 
the one-body Hermitian operator $\hat{F}$
\begin{equation}
 \hat{F}=\int\! \dd^3x\,\psi^\dagger(\vct{x})F\psi(\vct{x}),\label{f}
\end{equation}
with the Hamiltonian $H$, 
the sum is equal to its expectation value of the Hartree or Hartree-Fock 
ground state\cite{rowe,th}, 
\begin{equation}
\sum_{n}\omega_n|\mtrix{n}{\hat{F}}{0}|^2
= \frac{1}{2}\mtrix{\hf}{[\,\hat{F}\,,\,[H,\hat{F}\,]\,]}{\hf}.\label{s}
\end{equation}
At first glance, however, it seems that the above non-relativistic rule is
not extended to the relativistic RPA.  If the operator $F$ has
the only coordinate-dependence,
the double commutator with the Dirac equation vanishes, 
\begin{equation}
[ \,F(\vct{x})\,,\,[\,h_0\,,\,F(\vct{x})\,]\,] = 0, \label{dirac}
\end{equation}
since $h_0$ contains the only linear derivative,
\begin{equation*}
h_0(x)=-\,i\vct{\alpha}\!\cdot\!\vct{\nabla}+\gamma_0M,
\end{equation*}
in contrast to the nonrelativistic kinetic energy.

In fact, it was a long-standing problem for the last more than 50 years
in relativistic local field theory   
that the energy-weighted sum 
could not be expressed with the double commutator\cite{goto,schwinger,itz}. 
Recently, this paradox has been solved by the present authors\cite{ks5}. 
The expression with commutator itself is correct, but the commutator 
should not be calculated according to the usual way in the local field theory,
where the anticommutator Eq.(\ref{anti_com}) is employed.
Eq.(\ref{anti_com}) holds in the infinite momentum space.
The commutator should be calculated first in a finite momentum space.
Next, its ground-state expectation value is evaluated, and then   
we make the momentum space infinite to obtain the positive sum value. 

In other words, in the local field theory, the commutator between the time
and space component of the nuclear four-current disappears,  
yielding the sum value to vanish. 
The commutator calculated in the finite momentum space, however, 
does not vanish, and its ground-state expectation value exists
even in the infinite momentum space.

Keeping this new insight in mind, 
let us investigate whether or not the same relationship 
as in Eq.(\ref{s}) holds in the relativistic RPA.
We will first derive an expression as to the double commutator
of the one-body operator with the total Hamiltonian,
and next calculate the sum of the energy-weighted strengths
for exciting the RPA states, in the same way as in the nonrelativistic
RPA \cite{rowe, th}.

The one-body operators,
$\hat{P}$ and $\hat{Q}$, are written as
\[
 \hat{P}=\int\!\dd^3x\,\psi^\dagger(\vct{x})P\psi(\vct{x})\,,\qquad
 \hat{Q}=\int\!\dd^3x\,\psi^\dagger(\vct{x})Q\psi(\vct{x}).
\]
Here, it is not  necessary for $P$ and $Q$ to be a function of the
only coordinate.
The Hamiltonian in the $\sigma$-$\omega$ model is given by
\[
 H=H_0+H_{\mathrm{int}},
\]
where each term is described as
\begin{align}
  H_0 &= \int\!\dd^3x\,
 \psi^\dagger(\vct{x})h_0\psi(\vct{x}),\\
 H_{\mathrm{int}}&=
 \frac12\int\!\dd^3x_1\,\dd^3x_2\,
 \psi^\dagger(\vct{x}_1)\psi^\dagger(\vct{x}_2)\calV_{12}
 \psi(\vct{x}_2)\psi(\vct{x}_1).
\end{align}

When we employ Eq.(\ref{anti_com}) as in  the local field theory,
the expectation value of the double commutator
with the one-body Hamiltonian $H_0$ is expressed as 
\begin{equation}
 \mtrix{\hf}
{\bigl[\,\hat{P}\,,\,
 \bigl[\,H_0\,,\,\hat{Q}\,\bigr]\,\bigr]}{\hf}
=
\sum_\alpha \theta_\alpha
\mtrix{\alpha}
{\bigl[\,P\,,\,
 \bigl[\,h_0\,,\,Q\,\bigr]\,\bigr]
}{\alpha}.\label{com_h0}
\end{equation}
As mentioned before, 
if the operator $P$ and $Q$  depend on the only coordinate,
the above double commutator vanishes. 
For a while, however, we keep the above form, and will be 
discussed later how the commutator should be calculated.

The double commutator with the two-body interaction is composed
of the two parts,
\[
 \mtrix{\hf}{
\bigl[\,\hat{P}\,,\,
\bigl[\,H_{\mathrm{int}}\,,\,\hat{Q}\,\bigr]\,
\bigr] 
}{\hf}
=
S_1+S_2\,,\qquad
S_k=
\sum_{\alpha\alpha'}\theta_\alpha\theta_{\alpha'}
\mtrix{\alpha,\alpha'}
{\bigl[\,P_k\,,\,\bigl[\,\calV_{12}\,,\,Q_1\,\bigr]\,\bigr]}
{\alpha,\alpha'},
\]
where we have neglected the exchange term, and the suffixes of $P$ and $Q$ 
correspond to those of the two-body interaction.
Employing, for example, the equation
\begin{align}
 \sum_{\alpha\alpha'}\theta_\alpha\theta_{\alpha'}
 \mtrix{\alpha,\alpha'}{P_1\calV_{12}Q_1}{\alpha,\alpha'}
 &=
 \sum_{\alpha\beta\beta'}
 \theta_\alpha
 \mtrix{\alpha}{P}{\beta}
 \mtrix{\beta}{V_{\mathrm{H}}}{\beta'}
 \mtrix{\beta'}{Q}{\alpha}\nonumber\\
 &=
 \sum_{\alpha}
 \theta_\alpha
 \mtrix{\alpha}{P V_{\mathrm{H}}Q}{\alpha},
\end{align}
derived from the closure property in the intermediate states,
the above $S_1$ is described in terms of the Hartree potential of
Eq.(\ref{hpot}) ,
\begin{equation}
 S_1=
\sum_{\alpha}\theta_\alpha
\mtrix{\alpha}
{\bigl[\,P\,,\,\bigl[\,V_{\mathrm{H}}\,,\,Q\,\bigr]\,\bigr]}
{\alpha}. \label{dbl_com_hartree_v}
\end{equation}
The similar calculation gives
\begin{eqnarray}
S_2=
\sum_{\alpha\alpha'\beta\beta'}
\left(\theta_\beta-\theta_\alpha\right)
\left(\theta_{\beta'}-\theta_{\alpha'}\right)
P_{\beta\alpha}Q_{\alpha'\beta'}V_{\alpha\beta',\beta\alpha'}.
\label{com_two}
\end{eqnarray}
Finally, Eqs.(\ref{com_h0}), (\ref{dbl_com_hartree_v})
and (\ref{com_two}) lead to
\begin{eqnarray}
\mtrix{\hf}{\bigl[\,\hat{P}\,,\,\bigl[\,H\,,\,\hat{Q}\,\bigr]\,\bigr] 
}{\hf}
= 
\sum_\beta \theta_\beta
\mtrix{\beta}
{
\bigl[\,P\,,\,
\bigl[
\,h\,,\,Q\,
\bigr]\,\bigr]
}{\beta}
+S_2,
\end{eqnarray}
where the Hartree Hamiltonian $h$ has been given in Eq.(\ref{hpot}).
If the first term is written, employing the closure property, as
\begin{align}
 \sum_\beta \theta_\beta
 \mtrix{\beta}
 {\bigl[\,P,\,\bigl[\,h\,,\,Q\, \bigr]\,\bigr]}{\beta}
&=
\sum_{\alpha\beta}(1-\theta_\alpha)\theta_\beta\Bigl(
 \mtrix{\beta}{P}{\alpha}
\mtrix{\alpha}{\bigl[\,h\,,\,Q\,\bigr]}{\beta} \Bigr. \nonumber \\
&
\Bigl.
\hphantom{\sum(1-\theta_\alpha)\theta_\beta\Bigl(} 
-\mtrix{\beta}{\bigl[\,h\,,\,Q\,\bigr]}{\alpha}
 \mtrix{\alpha}{P}{\beta}
 \Bigr)
\label{equal}
\end{align}
we obtain the expression
\begin{align}
 \mtrix{\hf}{
 \bigl[\,\hat{P}\,,\,
 \bigl[\,H\,,\,\hat{Q}\,\bigr]\,\bigr] 
 }{\hf}
&=
 \sum_{\alpha\beta}\left( \theta_\beta-\theta_\alpha\right)
 E_{\alpha\beta}P_{\beta\alpha}Q_{\alpha\beta} \nonumber \\
&
 +
 \sum_{\alpha\alpha'\beta\beta'}
 \left(\theta_\beta-\theta_\alpha\right)
 \left(\theta_{\beta'}-\theta_{\alpha'}\right)
 P_{\beta\alpha}Q_{\alpha'\beta'}V_{\alpha\beta',\beta\alpha'}.
\end{align}
Using the notation $N_{\alpha\beta}$, instead of the step function,
it is described as,
\begin{align}
 \mtrix{\hf}{
\bigl[\,\hat{P}\,,\,\bigl[\,H\,,\,\hat{Q}\,\bigr]\,\bigr] 
 }{\hf}
 &=
 \sum_{\alpha\beta}\left( N_{\alpha\beta}-N_{\beta\alpha}\right)
 E_{\alpha\beta}P_{\beta\alpha}Q_{\alpha\beta}\nonumber \\
 &\  +
 \sum_{\alpha\alpha'\beta\beta'}
 \left( N_{\alpha\beta}-N_{\beta\alpha}\right)
 \left( N_{\alpha'\beta'}-N_{\beta'\alpha'}\right)
 P_{\beta\alpha}Q_{\alpha'\beta'}V_{\alpha\beta',\beta\alpha'},
 \label{double_com_value}  
\end{align}
which is also written with the notation Eq.(\ref{def_rrpa_ab}) as
\begin{equation}
\mtrix{\hf}{
\bigl[\,\hat{P}\,,\,
\bigl[\,H\,,\,\hat{Q}\,\bigr]
\bigr] 
}{\hf}
=
\sum_{\alpha\alpha'\beta\beta'}
\bigl(\,P_{\beta\alpha}\ -\!P_{\alpha\beta}\,\bigr)
\left(
\begin{array}{cc}
A_{\alpha\beta,\alpha'\beta'} & B_{\alpha\beta,\alpha'\beta'} \\[2pt]
B^\ast_{\alpha\beta,\alpha'\beta'} & A^\ast_{\alpha\beta,\alpha'\beta'} \\
\end{array}
\right)
\left(
\begin{array}{c}
Q_{\alpha'\beta'} \\[2pt]
-\,Q_{\beta'\alpha'}
\end{array}
\right).
\label{double_com_value_matrix}
\end{equation}

In the No$\nbar$ approximation, $\hat{P}$, $\hat{Q}$ and 
$H$  
in the above equations should be replaced by 
\begin{equation}
\hat{P}_+=\int\!\dd^3x\,
\psi^\dagger(\vct{x})\varLambda P\varLambda\,\psi(\vct{x})\,,\quad
\hat{Q}_+=\int\!\dd^3x\,
\psi^\dagger(\vct{x})\varLambda Q\varLambda\,\psi(\vct{x})\,,
 \label{nonbar1}
\end{equation}
and 
\begin{align}
 H_+ &=  (H_0)_+ + (H_{\mathrm{int}})_+ \,, \label{nonbar2}\\
 &  (H_0)_+ = \int\!\dd^3x\,
 \psi^\dagger(\vct{x})\varLambda h_0\varLambda\,\psi(\vct{x})\,,
 \nonumber\\[4pt]
 &  (H_{\mathrm{int}})_+
 =\frac12\int\!\dd^3x_1\,\dd^3x_2\,
 \psi^\dagger(\vct{x}_1)\psi^\dagger(\vct{x}_2)
 \varLambda_1\varLambda_2\calV_{12}\varLambda_1\varLambda_2\,
 \psi(\vct{x}_2)\psi(\vct{x}_1),\nonumber
\end{align}
with the projection operator
$ \varLambda(\vct{x})=\sum\limits_{E_\alpha>0}\varphi_{\alpha}(\vct{x})
\varphi^\dagger_{\alpha}(\vct{x})$.

Next, we calculate directly the energy-weighted sum, 
\begin{equation}
 \srpa =\sum_{n(\lambda_n=1)} 
\omega_n\bigl| \mtrix{n}{\hat{F}}{0} \bigr|^2.
\end{equation}
In writing the matrix elements
\[
 \mtrix{n}{\hat{F}}{0}
=\sum_{\alpha\beta}F_{\alpha\beta}C^{(n)\,\ast}_{\alpha\beta}\,,\qquad
F_{\alpha\beta}=\mtrix{\alpha}{F}{\beta}
=\int\!\dd^3x\,\varphi_\alpha^\dagger(\vct{x})F\varphi_\beta(\vct{x}),
\]
the sum is expressed as
\begin{equation}
 \srpa = 
\sum_{\alpha\beta\alpha'\beta'}
F^\ast_{\alpha\beta}F_{\alpha'\beta'}
\sum_{n(\lambda_n=1)} 
\omega_n
C^{(n)}_{\alpha\beta}C^{(n)\,\ast}_{\alpha'\beta'}.
\end{equation}
From the relationship $F_{\alpha\beta}=F^\ast_{\beta\alpha}$
and
\[
 \sum_{n(\lambda_n=1)} 
\omega_n
C^{(n)}_{\alpha\beta}C^{(n)\,\ast}_{\alpha'\beta'}
=
\sum_{n(\lambda_n=-1)} 
(-\,\omega_n)
C^{(n)\,\ast}_{\beta\alpha}\,C^{(n)}_{\beta'\alpha'},
\]
we have
\begin{equation}
\srpa 
=
\frac12
\sum_{\alpha\beta\alpha'\beta'}
F^\ast_{\alpha\beta}F_{\alpha'\beta'}
\sum_{n}
\lambda_n 
\omega_n
C^{(n)}_{\alpha\beta}C^{(n)\,\ast}_{\alpha'\beta'}.
\end{equation}
Since Eqs.(\ref{eq_coef_c}) and (\ref{comp_cc}) provide us with
\begin{align*}
  \sum_{n}
 \lambda_n 
 \omega_n
 C^{(n)}_{\alpha\beta}C^{(n)\,\ast}_{\alpha'\beta'}
 &=
 \sum_{n}
 \lambda_n C_{\alpha'\beta'}^{(n)\,\ast}
 \Bigl(
 E_{\alpha\beta}C_{\alpha\beta}^{(n)}
 +(N_{\alpha\beta}-N_{\beta\alpha})\sum_{\alpha''\beta''}
 V_{\alpha\beta'',\beta\alpha''}C_{\alpha''\beta''}^{(n)}
 \Bigr) \\
 &=
 (N_{\alpha\beta}-N_{\beta\alpha})E_{\alpha\beta}
 \,\delta_{\alpha\alpha'}\delta_{\beta\beta'}
 +
 (N_{\alpha\beta}-N_{\beta\alpha})
 (N_{\alpha'\beta'}-N_{\beta'\alpha'})
 V_{\alpha\beta',\beta\alpha'},
\end{align*}
finally we obtain the RPA sum value
\begin{align}
 \srpa 
 &= \frac12\sum_{\alpha\beta} 
 (N_{\alpha\beta}-N_{\beta\alpha})E_{\alpha\beta}|F_{\alpha\beta}|^2\nonumber\\
 &\phantom{=} +
 \frac12
 \sum_{\alpha\beta}
 \sum_{\alpha'\beta'}
 F^\ast_{\alpha\beta}F_{\alpha'\beta'}
 (N_{\alpha\beta}-N_{\beta\alpha})
 (N_{\alpha'\beta'}-N_{\beta'\alpha'})
 V_{\alpha\beta',\beta\alpha'}.\label{sum_rpa}
\end{align}
The first term of the above equation is nothing but the sum value
in the Hartree approximation, $\sha$.
For each approximation, it is given by,
\begin{equation}
 \sha =\sum_{ph}E_{ph}|F_{ph}|^2 +
\left\{
\begin{array}{ll}
\sum\limits_{p\bar{n}}E_{p\bar{n}}|F_{p\bar{n}}|^2 \, & \mbox{(Full),}  \\[10pt]
\sum\limits_{h\bar{n}}E_{\bar{n}h}|F_{\bar{n}h}|^2 \, & \mbox{(NSA, NFA),} \\[10pt]
0 \, & (\mathrm{No}\nbar).\,
\end{array}
\right.
\label{hews}
\end{equation}
In the Full case, $\sha$ is positive, as should be so, but
in NSA and NFA, the second term of the right hand side, coming from
the $\nbar$ contributions,
is negative owing to $E_{\bar{n}h}=-E_{h\bar{n}}<0$.
In Ref.\cite{ks5}, it has been shown that the negative contributions
exactly cancel the first term from the particle-excitations
in nuclear matter, yielding $\sha=0$.
In writing the term in the Full approximation as
\begin{equation}
\sum_{p\bar{n}}E_{p\bar{n}}|F_{p\bar{n}}|^2
=\sum_{n\bar{n}}E_{n\bar{n}}|F_{n\bar{n}}|^2 
-\sum_{h\bar{n}}E_{h\bar{n}}|F_{\bar{n}h}|^2,
\end{equation}
$n$ indicating the nucleon states $(p+h)$, NSA and NFA neglect
the first term of the right hand side which makes the left hand side
positive.

Now comparing Eq.(\ref{double_com_value})
with Eq.(\ref{sum_rpa}), we obtain the relationship
\begin{equation}
 \srpa
=\frac12
\mtrix{\hf}{
\bigl[\,\hat{F}\,,\,
\bigl[\,H\,,\,\hat{F}\,\bigr]\,
\bigr] 
}{\hf}\,.\label{rpasumrule}
\end{equation}
Thus, we can express formally the energy-weighted sum
in terms of the double commutator,
as in non-relativistic RPA\cite{rowe,th}.
In the No$\nbar$ approximation, $\hat{F}$ and $H$
in the above equations should be replaced as in Eqs.(\ref{nonbar1})
and (\ref{nonbar2}).

The above double commutator, however, should not
be calculated using Eq.(\ref{anti_com}) in order to obtain the sum value,
$\srpa$, given by Eq.(\ref{sum_rpa}).
Instead of Eq.(\ref{anti_com}) defined in the infinite momentum space,
we must use the anti-commutation relation
in a finite momentum space\cite{ks5}, 
\begin{equation}
\{\psi_a(\vct{x})\,,\,\psi_b^\dagger(\vct{y})\,\}=
\delta_{ab}d(\vct{x}-\vct{y})  ,\label{newcom}
 \end{equation}
where $d(\vct{x})$ is defined as
\begin{equation}
d(\vct{x})
 =\int\!\frac{\dd^3p}{(2\pi)^3}\,\Theta_{\vct{p}}e^{i\vct{p}\cdot\vct{x}},
\quad  \Theta_{\vct{p}} =\theta(P_\infty-|\vct{p}|).
\end{equation}
Then, we have its ground-state expectation value which does not vanish. 
By making $P_\infty$ infinite in the expectation value,
we can obtain the sum value.
We note that Eq.(\ref{newcom}) is reduced to Eq.(\ref{anti_com})
in the limit $P_\infty \rightarrow \infty$, because of
\begin{equation}
d(\vct{x})
 \xrightarrow{P_\infty\rightarrow \infty\,}\delta(\vct{x}).
\end{equation}

The double commutator in the Hartree approximation has been calculated 
for nuclear matter, according to Eq.(\ref{newcom}) in Ref.\cite{ks5}.
In the Full space $N_{\alpha\beta}$,
the sum value of $\sha$ is divergent,
being proportional to $P^2_\infty$, in contrast to $\sha=0$
in NSA and NFA.

The energy-weighted sum for the Gamow-Teller transition strengths
in the relativistic RPA has been explored in Ref.\cite{ks3}.

\section{The continuity equation}\label{ce}

The current conservation in the relativistic RPA has been studied
in previous papers\cite{chin,ks1,ks2,srm}.
One of the reasons why NFA and NSA were accepted so far is
because they do not violate the current conservation. 
Generally speaking, it is important for phenomenological models
to keep at least well-known fundamental principles. In particular,
Refs.\cite{ks1, srm} investigated the electron scattering where
the continuity equation must be essential.

There are various ways to show the current conservation 
in NFA and NSA. One of the ways is to write the Hartree
polarization function in  Eq.(\ref{def_rpa_pola})
in terms of the Hartree propagator, $\grH(1,2)$,
\begin{equation}
i\piH(A(1),B(2))
=
\mathrm{Tr}\Bigl(
A\grH(1,2)B \grH(2,1)
\Bigr)
\label{def_pi_h},
\end{equation}
where $\grH(1,2)$ is given by\cite{ks1}
\begin{eqnarray}
  \grH(1,2) =
 \frac{1}{2\pi}
\int_{-\infty}^\infty\!\!\dd\omega\,e^{-i\omega(t_1-t_2)}\,
 \grH(\vct{x}_1,\vct{x}_2,\omega), \label{gh_expand_eigen2}
\end{eqnarray}
with
\begin{equation}
  \grH(\vct{x}_1,\vct{x}_2,\omega)
=
\sum_\alpha
\varphi_\alpha(\vct{x}_1)
\overline{\varphi}_\alpha(\vct{x}_2)
\left(
\frac{1-\theta_\alpha}{\omega-E_\alpha+i\varepsilon}
+\frac{\theta_\alpha}{\omega-E_\alpha-i\varepsilon}
\right).
\label{def_gh_x1_x2}
\end{equation}
Owing to the closure property Eq.(\ref{comp}),
the above Hartree propagator satisfies 
\begin{equation}
 \Bigl( i\gamma_\mu\partial_x^\mu -M-\varSigma(x)  \Bigr)\grH(x,y)
= \delta(x-y). \label{hp_delta}
\end{equation}
Then, Eq.(\ref{def_pi_h}) provides us with
\begin{align}
 \partial_x^\mu\piH(\gamma_\mu(x),B(y)) 
 &=
 -\,\mathrm{Tr}\!\left[
 \Bigl(
 \delta(x-y)+\bigl( M+\varSigma(x) \bigr)\grH(x,y)
 \Bigr)B\grH(y,x)
 \right] \nonumber \\
& \phantom{=\,} 
+ \mathrm{Tr}\!\left[
 \grH(x,y)B
 \Bigl(
 \delta(x-y)+\grH(y,x)\bigl(M+\varSigma(x)\bigr)
 \Bigr)
 \right]
 = 0. \label{hcurrent}
\end{align}
The above equation and Eq.(\ref{def_rpa_pola}) imply for
the RPA polarization function to satisfy
\begin{equation}
 \partial_x^\mu \varPi(\gamma_\mu(x),B(y))=0.\label{rpace}
\end{equation}
Thus, as far as Eq.(\ref{hp_delta}) holds, 
the relativistic RPA conserves the current.
As mentioned before, Eq.(\ref{hp_delta}) requires
the complete set of eigenfunctions for the Hartree field.
If $\nbar$ states are neglected, 
Eq.(\ref{hp_delta})
does not hold as,
\begin{equation}
 \Bigl( i\gamma_\mu\partial_x^\mu -M-\varSigma(x)  \Bigr)\grH(x,y)
=
\delta(x_0-y_0)
\sum_{\alpha}
\varphi_\alpha(\vct{x})
\varphi_\alpha^\dagger(\vct{y})
\left( p_\alpha+h_\alpha \right).\label{hp_ph}
\end{equation}

The above proof of Eq.(\ref{rpace}) seems to be independent of
how the Hartree ground state is occupied by N and $\nbar$.
Indeed, both NSA and NFA satisfy this equation.
Physically, however, $\nbar$ states should be occupied
in the ground state,
as far as the negative energy states are required in the model
framework.
Hence, it may be instructive to see explicitly
how the continuity equation is satisfied,
when only a part of N-$\nbar$ excitations
are included in the calculations of the excited states.

Using Eqs.(\ref{hpf}) to (\ref{nalphabta}), the Hartree polarization
function can be written as,
\begin{equation}
i \piH(\gamma_\mu(1),B(2)) 
= \sum_{\alpha\beta}F_{\mu}^{(\alpha\beta)}(\vct{x},\vct{y})
T^{(\alpha\beta)}(x_0,y_0),\label{pi0_spec}
\end{equation}
where we have used the notations:
\begin{align*}
 F_{\mu}^{(\alpha\beta)}(\vct{x},\vct{y})
 &=
 \overline{\varphi}_\beta(\vct{x})\gamma_\mu \varphi_\alpha(\vct{x})\,
 \overline{\varphi}_\alpha(\vct{y}) B \varphi_\beta(\vct{y}) \\[4pt]
 T^{(\alpha\beta)}(x_0,y_0)
 &= \Bigl(
 N_{\alpha\beta}\theta(x_0-y_0)
 +N_{\beta\alpha}\theta(y_0-x_0)
 \Bigr)\,e^{-iE_{\alpha\beta}(x_0-y_0)}.
\end{align*}
Then, we obtain
\begin{align}
  i\,
 \partial_x^\mu \piH(\gamma_\mu(x),B(y)) 
 &= 
 \delta(x_0-y_0)\sum_{\alpha\beta}
 F_{0}^{(\alpha\beta)}(\vct{x},\vct{y})
 \Bigl( N_{\alpha\beta}-N_{\beta\alpha} \Bigr) \nonumber \\
 & + 
 \,\sum_{\alpha\beta}
 \Bigl( 
 \partial_x^kF_{k}^{(\alpha\beta)}(\vct{x},\vct{y})
 -iE_{\alpha\beta}F_{0}^{(\alpha\beta)}(\vct{x},\vct{y})
 \Bigr)\,T^{(\alpha\beta)}(x_0,y_0).
 \label{eq_conv_c}
\end{align}
Since $\varphi_\alpha(\vct{x})$ is an eigenfunction of the
Hartree Hamiltonian, the second term of the right hand side vanishes.
Employing the expression of Eq.(\ref{nab-nba}) for the full space,
and then the closure relation Eq.(\ref{comp}), the above equation
can be written as
\begin{align}
i\,\partial_x^\mu \piH(\gamma_\mu(x),B(y))
&=
 \delta(x-y)\biggl(
 \sum_{\beta}
 \overline{\varphi}_\beta(\vct{x}) B \varphi_\beta(\vct{x})
 \, h_\beta
 -
 \sum_{\alpha}
 \overline{\varphi}_\alpha(\vct{x}) B \varphi_\alpha(\vct{x})
 \, h_\alpha\biggr. \nonumber\\
 & 
\biggl.
\hphantom{\delta(x-y)}
+ 
 \sum_{\beta}
 \overline{\varphi}_\beta(\vct{x}) B \varphi_\beta(\vct{x})
 \, \bar{n}_\beta
 -
 \sum_{\alpha}
 \overline{\varphi}_\alpha(\vct{x}) B \varphi_\alpha(\vct{x})
 \, \bar{n}_\alpha \biggr)
=0.
 \label{eq_conv_c_hartree}
\end{align}
This equation shows the first and the second line
vanishing separately. 
As a result, NSA and NFA satisfy the continuity equation,
although they have the only first line.
The second line comes from the divergent terms describing the 
excitations of $\nbar$ in the Dirac sea, 
$N_{\alpha\beta}=(p_\alpha+h_\alpha)\bar{n}_\beta$.
This fact implies that the vacuum should satisfy 
the continuity equation by itself, and that the current conservation is
independent of the divergence problem.
Thus even if the continuity equation is described correctly,
it is not assured that the same approximation is applicable to
calculations of other physical quantities.

We note finally that Eq.(\ref{rpace}) provides us with the familiar form
of the continuity equation to be expressed as the transition matrix
element,
\[
 \sum_{\alpha\beta}
\Bigl(
-\,i\omega_n J_{\beta\alpha}^0(\vct{x})
+\vct{\nabla}\!\cdot\!\vct{J}_{\beta\alpha}(\vct{x})
\Bigr)
C_{\alpha\beta}^{(n)}
=0,\qquad
J^\mu_{\beta\alpha}(\vct{x})
=\overline{\varphi}_\beta(\vct{x})\gamma^\mu \varphi_\alpha(\vct{x}).
\]

\section{The spurious state}\label{ss}

The present relativistic RPA equation has spurious solutions
in the same way as in non-relativistic models\cite{rowe,th}.
If $[\,H\,,\,\hat{Q}\,]=0$, Eq.(\ref{double_com_value_matrix})
provides us with
\[
\sum_{\alpha\alpha'\beta\beta'}
\bigl(\,P_{\beta\alpha}\ -\!P_{\alpha\beta}\,\bigr)
\left(
\begin{array}{cc}
A_{\alpha\beta,\alpha'\beta'} & B_{\alpha\beta,\alpha'\beta'} \\[2pt]
B^\ast_{\alpha\beta,\alpha'\beta'} & A^\ast_{\alpha\beta,\alpha'\beta'} \\
\end{array}
\right)
\left(
\begin{array}{c}
Q_{\alpha'\beta'} \\[2pt]
-\,Q_{\beta'\alpha'}
\end{array}
\right)
=0.
\]
The above equation holds for any $P_{\alpha\beta}$, so that we have
\begin{equation}
 \sum_{\alpha'\beta'}
\left(
\begin{array}{cc}
A_{\alpha\beta,\alpha'\beta'} &
B_{\alpha\beta,\alpha'\beta'} \\[2pt]
B_{\alpha\beta,\alpha'\beta'}^\ast &
A_{\alpha\beta,\alpha'\beta'}^\ast
\end{array}
\right) 
\left(
\begin{array}{c}
Q_{\alpha'\beta'} \\[2pt]
-\,Q_{\beta'\alpha'}
\end{array}
\right)
=0. \label{spurious_mode}
\end{equation}
Thus, when $Q_{\alpha\beta}\ne0$, $\hat{Q}$ is the spurious solution
of Eq.(\ref{rel_rpa_eq}) with $\omega_n=0$.
If the Full RPA has the spurious state, NSA and NFA also separate it from
the other solutions. 
As seen in Eq.(\ref{def_rrpa_ab}), $A$ and $B$ depend on $N_{\alpha\beta}$.
Therefore, the spurious states can not be described without $\nbar$-degrees
of freedom not only in the Full calculation, but also in NSA and NFA\cite{df}.
In the No$\nbar$ approximation,
$[\,H_+\,,\,\hat{Q}_+\,]=0$ would be required
for $\hat{Q}_+$ to be the spurious solution, but not be assured by
$[\,H\,,\,\hat{Q}\,]=0$.

Up to this stage, it seems that the spurious state is well described
in NSA and NFA.
As discussed in \S\ref{r}, however, unphysical effects coming from
the disregard of the divergent terms are hidden in their RPA equation.
In order to see this fact in more detail,
let us describe Eq.(\ref{spurious_mode}) explicitly.
On the one hand, it is written in the Full case as
\begin{equation}
E_{\alpha\beta} Q_{\alpha\beta}
+
\sum_{\alpha'\beta'}(p_{\alpha'} h_{\beta'}+p_{\alpha'}\bar{n}_{\beta'})
\Bigl(V_{\alpha\beta',\beta\alpha'}
Q_{\alpha'\beta'}
-
V_{\alpha\alpha',\beta\beta'}
Q_{\beta'\alpha'}\Bigr)
=0.\label{fulls}
\end{equation}
On the other hand, it becomes in NFA,
\begin{equation}
E_{\alpha\beta} Q_{\alpha\beta}
+
\sum_{\alpha'\beta'}(p_{\alpha'} h_{\beta'}-h_{\alpha'}\bar{n}_{\beta'})
\Bigl(V_{\alpha\beta',\beta\alpha'}
Q_{\alpha'\beta'}
-
V_{\alpha\alpha',\beta\beta'}
Q_{\beta'\alpha'}\Bigr)
=0,\label{nfas}
\end{equation}
and in NSA, 
\begin{equation}
E_{\alpha\beta} Q_{\alpha\beta}
+
\sum_{\alpha'\beta'}(p_{\alpha'} h_{\beta'}+\bar{n}_{\alpha'} h_{\beta'})
\Bigl(V_{\alpha\beta',\beta\alpha'}
Q_{\alpha'\beta'}
-
V_{\alpha\alpha',\beta\beta'}
Q_{\beta'\alpha'}\Bigr)
=0.\label{nsas}
\end{equation}
It is seen that Eq.(\ref{nsas}) is the same as Eq.(\ref{nfas}),
by exchanging the suffixes $\alpha'$ and $\beta'$ in the factor with
$\bar{n}_{\alpha'} h_{\beta'}$.
In writing $p_\alpha h_\beta+p_\alpha\bar{n}_\beta
=p_\alpha h_\beta - h_\alpha \bar{n}_\beta +(p_\alpha+h_\alpha)\bar{n}_\beta$
for the Full case, we can recognize in Eq.(\ref{nfas})
that NFA ignores the last term $(p_\alpha+h_\alpha)\bar{n}_\beta$,
so that the minus sign of
$ - h_\alpha \bar{n}_\beta$ changes unphysically
the one of the interactions in NFA and NSA.

Let us give a few comments on the above discussions.
Eq.(\ref{spurious_mode}) for Full and NSA
can be also derived by calculating the matrix element,  
\begin{equation}
\mtrix{\hf}{a^\dagger_\beta a_\alpha [H, Q]}{\hf}=0\label{hq0}
\end{equation}
for $[H, Q] = 0$.
Remembering the Hartree grand state $\ket{\hf}$ to be different
from each other in Full and NSA, the above equation is expressed
using $N_{\alpha\beta}$ as
\begin{equation}
N_{\alpha\beta}\sum_{\alpha'\beta'}Q_{\alpha'\beta'}
\Bigl(
N_{\alpha'\beta'}
\mtrix{\hf}{a_\beta^\dagger a_\alpha H a_{\alpha'}^\dagger a_{\beta'}}{\hf}
- N_{\beta'\alpha'}
\mtrix{\hf}{a_\beta^\dagger a_\alpha a_{\alpha'}^\dagger a_{\beta' }H}{\hf}\Bigr)
=0,\label{hq}
\end{equation}
which leads to, neglecting the exchange terms, 
\begin{equation}
\sum_{\alpha'\beta'}\Bigl(A_{\alpha\beta,\alpha'\beta'}Q_{\alpha'\beta'}
 -B_{\alpha\beta,\alpha'\beta'}Q_{\beta'\alpha'}\Bigr)=0.
\end{equation}
Combining this with a similar equation from $[H, Q^\dagger]=0$, we obtain 
Eq.(\ref{spurious_mode}).

Notice that $N_{\alpha\beta}$ in Eq.(\ref{hq}) is not simply replaced 
by the one of NFA. The step from  Eq.(\ref{hq0}) to Eq.(\ref{hq}) 
requires the complete set for
$N_{\alpha'\beta'}$, whereas NFA ignores the vacuum polarization part,
$(p_\alpha+h_\alpha)\bar{n}_\beta$, of
$p_\alpha h_\beta+p_\alpha\bar{n}_\beta
=p_\alpha h_\beta - h_\alpha \bar{n}_\beta
+(p_\alpha+h_\alpha)\bar{n}_\beta$
in the Full case.
Indeed, the vacuum polarization part of Eq.(\ref{fulls})
does not satisfy 
\begin{equation}
\sum_{\alpha'\beta'}(p_{\alpha'}+h_{\alpha'}) \bar{n}_{\beta'}
\Bigl(V_{\alpha\beta',\beta\alpha'}
Q_{\alpha'\beta'}
-
V_{\alpha\alpha',\beta\beta'}
Q_{\beta'\alpha'}\Bigr)
= 0.
\end{equation}
and the terms corresponding to Eq.(\ref{nfas}) in Eq.(\ref{fulls}) also
do not vanish.
When Eq.(\ref{nsas}) for NSA holds, however,
Eq.(\ref{nfas}) for NFA also does.
This fact implies that neglecting the vacuum polarization in RPA 
is equivalent to assuming empty $\nbar$ states
as in NSA with the complete set for $N_{\alpha\beta}$. 
In order to render Eq.(\ref{nfas}) valid in NFA,
as well as Eq.(\ref{nsas}) in NSA, 
the nuclear interactions for the Full case have to be modified
self-consistently together with $E_{\alpha\beta}$.
Even so, as in the case of the continuity equation,
the description of the spurious state in NFA and
NSA does not provide us with a justification of their approximations.

In the No$\nbar$ approximation,
$[H, Q]$ in Eq.(\ref{hq0}) may be replaced by the projected one,
$[H_+, Q_+] =0$, in addition to the Hartree ground state.

\section{Schematic model}\label{sm}

It is well-known in non-relativistic RPA that a schematic model
with separable interactions illustrates well the general character
of the RPA solutions\cite{rowe}.
Let us explore the structure of NFA, NSA and the full relativistic RPA
by using a similar model.

 We assume the interaction $V$ of which matrix elements are given by
\begin{equation}
  V_{\alpha\beta',\beta\alpha'}
=\sum_a \kappa_a f^{(a)}_{\alpha\beta}f^{(a)\,\ast}_{\alpha'\beta'}.
\label{int_separable}
\end{equation}
Then,
Eq.(\ref{eq_coef_c}) provides us with
\[
 C_{\alpha\beta}=\frac{N_{\alpha\beta}-N_{\beta\alpha}}{\omega-E_{\alpha\beta}}
\sum_a f_{\alpha\beta}^{(a)}{\cal N}_a\,,\qquad
{\cal N}_a
=\kappa_a\sum_{\alpha\beta}f^{(a)\ast}_{\alpha\beta} C_{\alpha\beta},
\]
which lead to
\[
 X_{\alpha\beta}=\frac{1}{\omega-E_{\alpha\beta}}
\sum_a f^{(a)}_{\alpha\beta}{\cal N}_a
\,,\qquad
 Y_{\alpha\beta}=-\,\frac{1}{\omega+E_{\alpha\beta}}
\sum_a f^{(a)}_{\beta\alpha}{\cal N}_a.
\]
Since we have
\begin{equation}
  {\cal N}_a=\kappa_a\sum_b R_{ab}(\omega){\cal N}_b\,,\qquad
R_{ab}(\omega)
=\sum_{\alpha\beta}
\frac{N_{\alpha\beta}-N_{\beta\alpha}}{\omega-E_{\alpha\beta}}
f^{(a)\ast}_{\alpha\beta}f^{(b)}_{\alpha\beta}
=R_{ba}^\ast(\omega),
\label{rpa_eigen_separable0}
\end{equation}
the eigenvalues of the RPA equation are determined by the
dispersion equation
\begin{equation}
  \mathrm{det}\Bigl(\delta_{ab}-\kappa_a R_{ab}(\omega)\Bigr)=0.
\label{rpa_eigen_separable}
\end{equation}
The normalization Eq.(\ref{oth_coef_c1}) becomes
\[
 \lambda=\sum_{\alpha\beta}
\bigl( N_{\alpha\beta}-N_{\beta\alpha} \bigr)
|C_{\alpha\beta}|^2
=-\sum_{ab}
{\cal N}_a^\ast\frac{dR_{ab}}{d\omega}{\cal N}_b.
\]

In the Full RPA, $R_{ab}(\omega)$ is given by
\begin{align}
  R^{\rFull}_{ab}(\omega)
 &=
 \sum_{ph}
 \left(
 \frac{f^{(a)\ast}_{ph} f^{(b)}_{ph}}{\omega-E_{ph}}
 -
 \frac{f^{(a)\ast}_{hp} f^{(b)}_{hp}}{\omega+E_{ph}}
 \right)
 +
 \sum_{p\bar{n}}
 \left(
 \frac{f^{(a)\ast}_{p\bar{n}} f^{(b)}_{p\bar{n}}}
 {\omega-E_{p\bar{n}}}
 -
 \frac{f^{(a)\ast}_{\bar{n}p} f^{(b)}_{\bar{n}p}}
 {\omega+E_{p\bar{n}}}
 \right)\nonumber\\
 &= 
\rph_{ab}(\omega) +\rpn_{ab}(\omega).
\end{align}
The part $\rpn_{ab}$ for particle-$\nbar$ excitations can be written
by using Pauli blocking terms as
\begin{equation}
  \rpn_{ab}(\omega)
=\rnn_{ab}(\omega)-\rhn_{ab}(\omega)\,,
\end{equation}
with
\begin{equation}
\rnn_{ab}(\omega)
=
\sum_{n\bar{n}}
\left(
\frac{f^{(a)\ast}_{n\bar{n}} f^{(b)}_{n\bar{n}}}
{\omega-E_{n\bar{n}}}
-
\frac{f^{(a)\ast}_{\bar{n}n} f^{(b)}_{\bar{n}}}
{\omega+E_{n\bar{n}}}
\right), \quad
\rhn_{ab}(\omega)=
\sum_{\bar{n}h}
\left(
\frac{f^{(a)\ast}_{h\bar{n}} f^{(b)}_{h\bar{n}}}
{\omega-E_{h\bar{n}}}
-
\frac{f^{(a)\ast}_{\bar{n}h} f^{(b)}_{\bar{n}h}}
{\omega+E_{h\bar{n}}}
\right).
\label{def_rnn}
\end{equation}
Then, we may write
\begin{equation}
 R^{\rFull}_{ab}(\omega)=\rph_{ab}(\omega)-\rhn_{ab}(\omega)
 +\rnn_{ab}(\omega).\label{full_dis}
 \end{equation}
In the case of NSA and NFA, ($N_{\alpha\beta}-N_{\beta\alpha}$) 
is given by ($h_\beta-h_\alpha$) in $R_{ab}(\omega)$ of
Eq.(\ref{rpa_eigen_separable0}),
so that we have
\begin{equation}
  R^{\rNSA,\rNFA}_{ab}(\omega)
=\rph_{ab}(\omega)
-\rhn_{ab}(\omega).
\label{def_rhn}
\end{equation}
It is seen that NSA and NSA neglect $\rnn_{ab}(\omega)$ in the Full
expression Eq.(\ref{full_dis}), leaving the Pauli blocking term
$\rhn_{ab}(\omega)$ with the minus sign.

When the interaction has the only one component 
$V_{\alpha\beta',\beta\alpha'}=\kappa f_{\alpha\beta}f^\ast_{\alpha'\beta'}$,
we have simply
\begin{equation}
  R(\omega)=\sum_{\alpha\beta}
N_{\alpha\beta}
\left(
\frac{|f_{\alpha\beta}|^2}{\omega-E_{\alpha\beta}}
-\frac{|f_{\beta\alpha}|^2}{\omega+E_{\alpha\beta}}
\right),\qquad
\lambda
=-\,|{\cal N}|^2\frac{dR}{d\omega}\,,\label{one_comp}
\end{equation}
and the eigenvalues of the RPA excited states are obtained from the equation
\begin{equation}
1-\kappa R(\omega)=0,\qquad  
 \frac{dR}{d\omega}<0.  \label{dis_single}
\end{equation}
Eq.(\ref{one_comp}) becomes for the full space, and for NSA and NSA,
respectively, 
\begin{align}
 R_{\rFull}(\omega) 
&=
 \sum_{ph}
 \frac{2E_{ph}|f_{ph}|^2}{\omega^2-E_{ph}^2}
 +
 \sum_{p\bar{n}}
 \frac{2E_{p\bar{n}}|f_{p\bar{n}}|^2}{\omega^2-E_{p\bar{n}}^2}
 =\rph(\omega)
 +\rpn(\omega)\,, \label{dis1}\\
 R_{\rNSA,\rNFA}(\omega) 
&=
 \sum_{ph}
 \frac{2E_{ph}|f_{ph}|^2}{\omega^2-E_{ph}^2}
 -
 \sum_{\bar{n}h}
 \frac{2E_{h\bar{n}}|f_{h\bar{n}}|^2}{\omega^2-E_{h\bar{n}}^2}
 =\rph(\omega)
 -\rhn(\omega).\label{dis2}
\end{align}
As mentioned before, the sign of the second term
of the right hand side for NSA and NFA is
opposite to the one for the Full case. 
This changes the signs of $\kappa$ and derivative
in the part of the N-$\nbar$ excitations in 
Eq.(\ref{dis_single}). 
The former change makes the attractive force(repulsive) work as 
a repulsive(attractive) one, and the latter change produces
RPA states with negative excitation energies.

The above unphysical effects in NSA and NFA
are not independent of low lying RPA states.
When $\omega\ll 2M^\ast$,
$M^\ast$ being the nucleon effective mass,
on the one hand,
we can set the following equation into Eq.(\ref{dis1}) for the Full case,
\begin{equation}
\rpn(\omega)\approx \rpn(0)=
-\sum_{p\bar{n}}
 \frac{2|f_{p\bar{n}}|^2}{E_{p\bar{n}}}
=\rnn(0)+\rhn(0)<0 \,\ .
\end{equation}
Hence, the dispersion equation of Eq.(\ref{dis_single}) can be written as 
\begin{equation}
1-\kappa^{\rFull}_{\mathrm{eff}} \rph(\omega)\approx0
\,,\qquad
\kappa^{\rFull}_{\mathrm{eff}}
=\frac{\kappa}{1+\kappa|\rpn(0)|}\, .\label{dis_single2}
\end{equation}
On the other hand, in Eq.(\ref{dis2}) for NSA and NFA, we may use
an approximation,
\begin{equation}
\rhn(\omega)\approx \rhn(0)=
-\sum_{\bar{n}h}
 \frac{|2f_{h\bar{n}}|^2}{E_{h\bar{n}}}<0,
\end{equation}
so that we have the dispersion equation with 
$\kappa^{\rNSA,\rNFA}_{\mathrm{eff}}$,
instead of $\kappa^{\rFull}_{\mathrm{eff}}$ in Eq.(\ref{dis_single2}),
\begin{equation}
\kappa^{\rNSA,\rNFA}_{\mathrm{eff}}
=\frac{\kappa}{1-\kappa|\rhn(0)|}\, .\label{NSA_kappa}
\end{equation}
In the degenerate limit where all the particle-hole energies
$E_{ph}$ are put equal to $\epsilon$, the dispersion equation provides all the 
solutions but one to be trapped at the unperturbed energy, and the one at
\begin{equation}
\omega^2=\epsilon^2+2\kappa_{\mathrm{eff}}\epsilon\sum_{ph}|f_{ph}|^2,
\end{equation}
where $\kappa_{\mathrm{eff}}$ is given by $\kappa^{\rFull}_{\mathrm{eff}}$, or
$\kappa^{\rNSA,\rNFA}_{\mathrm{eff}}$.

The difference between the full space RPA, and NSA and NFA appears
in the denominator of $\kappa_{\mathrm{eff}}$.
The contribution from the coupling
with N-$\nbar$ states have an opposite sign to each other.
If the interaction is attractive(repulsive), $\kappa<0(>0)$, one should have 
$\kappa_{\mathrm{eff}}<\kappa$, which enhances(diminishes)
the attractive(repulsive) force. This fact is realized
in  Eq.(\ref{dis_single2}) for the full space RPA, whereas not in
Eq.(\ref{NSA_kappa}) for the NSA and NFA.

In fact, these effects of the Pauli blocking terms in NFA and NSA
were recognized in the  previous numerical calculations.
In the response functions to quasielastic electron scattering,
the effects are rather negligible\cite{ks1},
but not in the description of the low lying states like the spurious
state\cite{df} and giant monopole states\cite{piek,mgwvr,mcg}.

The RPA spurious state to be predicted at zero energy is dominated by 
attractive forces.
In relativistic models, the phenomenological attractive force is rather 
strong. Hence, in some case an imaginary solution of the RPA
equation was obtained, 
when the coupling with N-$\nbar$ states was ignored.
By taking into account the coupling, the attractive interaction is
weakened effectively, and reproduced the spurious state at zero energy\cite{df}.

The same thing happened in the calculation of
giant monopole states which were also sensitive to attractive forces.
Without the coupling, the calculated excitation energy was too low, but
the coupling played a role of the repulsive force,
and explained well the experimental values\cite{piek,mgwvr,mcg}.
Thus, 
the agreement with experimental
data does not mean always models to be physically proper
for description of phenomena.

More intuitive understanding of the difference between the full calculation,
and NSA and NFA is shown in the Fig 1 and 2.
Fig.1 shows schematically the dispersion relation of the full RPA, 
Eq.(\ref{dis1}), in the case of $\kappa<0$. It has a familiar structure
to be found in literature on non-relativistic RPA\cite{rowe}.
The thick and thin solid curves are calculated with and without the
coupling with the N-$\nbar$ states, respectively. The solid
circles denote the eigenvalues corresponding to $\lambda=1 (dR/d\omega<0)$,
while the open circles $\lambda=-1$.
It is seen that the excitation energy of the lowest state
is pushed down owing to the coupling, as expected.


\begin{figure}[ht]
\centering\includegraphics{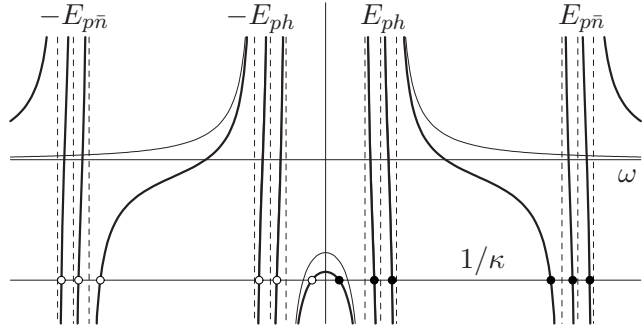}%
\caption{
Graphical solution of the full  RPA dispersion equation with
Eq.(\ref{dis1}). The thick and thin solid curves are calculated with
and without the coupling with the N-$\nbar$ states, respectively.
The solid circles denote the eigenvalues of the RPA excited states.
The vertical broken lines indicate the position of the unperturbed
energies. For the details, see the text.
}
\end{figure}

Fig.2 shows the dispersion relation with Eq.(\ref{dis2}) of NSA and NFA.
On the contrary to Fig.1, the lowest positive energy state is pushed up
with the coupling, in spite of the fact that the interaction is assumed 
to be attractive.
Moreover, except for low lying positive energy states around $E_{ph}$,
the eigenvalues with $\lambda=1 (dR/d\omega<0)$ appear
at the negative energy region
around $-E_{h\bar{n}}$ in both NSA and NFA.

In addition to the thick and thin curves as in Fig.1,
the dotted curve is shown, which is obtained without the coupling
and with a less number of particle-hole states. Compared with
the thin curve, we see that the excitation energy of the lowest state
is decreased with the increased number of the particle-hole states.
Therefore, when the attractive force is enough for the spurious state
with zero energy in NSA and NFA, its eigenvalue becomes imaginary
in neglecting the coupling with N-$\nbar$ states,
because of $\kappa^{-1}=\rph(0)-\rhn(0)>\rph(0)$ as in Eq.(\ref{dis2}).  
These dependence on the number of the configurations 
was also observed numerically in Ref.\cite{df}.


\begin{figure}[ht]
\centering\includegraphics{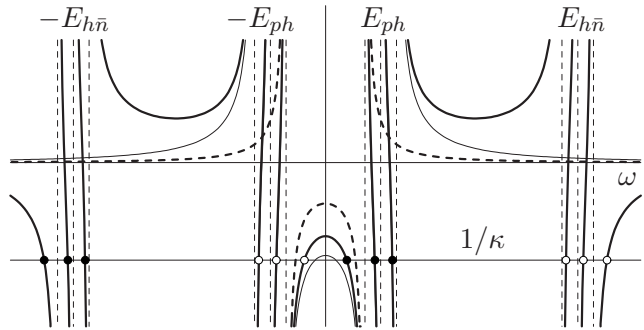}%
\caption{
Same as Fig.1 for Eq.(\ref{dis2}).
The dotted curve is obtained without the N-$\nbar$ coupling and
with a less number of particle-hole states. 
For the details, see the text.
}
\end{figure}

Finally, it may be useful to re-examine in the present RPA framework
the relativistic Landau-Migdal parameters of the $\sigma$-$\omega$ model 
developed in Refs.\cite{ks2,ks6,matsui}.
In discussing the response of nuclear matter at low momentum
transfer $q \approx 0$, the interaction Eq.(\ref{def_two_body_int})
of the $\sigma$-$\omega$ model can be expressed as a separable form. 
It is written in Eq.(\ref{int_separable}) as
\[
f^{(a)}_{\alpha\beta}
=
\frac{1}{\sqrt{\Omega}}
\mtrix{\overline{\alpha}}{e^{i\vct{q}\cdot\vct{x}}\varGamma^a}{\beta}
=\frac{\delta_{\vct{p}_\alpha,\vct{p}_\beta+\vct{q}}}{\sqrt{\Omega}}\,
\overline{w}_\alpha \varGamma^a w_\beta\,,\quad
\]
with
\[
\kappa_a
=
\left\{
\begin{array}{ll}
-\,f_\sigma=-\,g_\sigma^2/m_\sigma^2\,, & a=-\,1, \\[4pt]
+\,f_\omega=g_\omega^2/m_\omega^2\,, & a=0, \\[4pt]
-\,f_\omega\,, & a=1,\,2,\,3.
\end{array}
\right.
\]
Here, $\Omega$ denotes
the volume of the system which we need for rewriting the integral
of Eqs.(\ref{d_sigma2}) and (\ref{d_omega2}) with the summation, 
and $w_\alpha=w_s(\vct{p}\sigma)$ stands for the four-component
spinor, $\alpha$ representing $\{s=\pm, \vct{p}, \sigma\}$,
\begin{align}
  w_+(\vct{p}\sigma) 
 &= 
 \sqrt{ \frac{E_{\vct{p}}+M^\ast}{2E_{\vct{p}}}} 
 \left( 
 \begin{array}{c} 
 1 \\[4pt] 
 \dfrac{\vct{\sigma}\!\cdot\!\vct{p}}{E_{\vct{p}}+M^\ast} 
 \end{array} 
 \right) \,\chi_\sigma\, , \\[4pt] 
 w_-(\vct{p}\sigma) 
 &= 
 \sqrt{ \frac{E_{\vct{p}}+M^\ast }{2E_{\vct{p}}}} 
 \left( 
 \begin{array}{c} 
 -\dfrac{\vct{\sigma}\!\cdot\!\vct{p}}{E_{\vct{p}}+M^\ast} 
 \\[8pt] 
 1 
 \end{array} 
 \right)\,\chi_\sigma\, , 
\end{align}
with
$E_{\vct{p}} = \sqrt{\vct{p}^2+M^{\ast 2}}$ and 
the 2-component spinor, $\chi_{\sigma}$.

The matrix $R$ in Eq.(\ref{rpa_eigen_separable}) can be divided into
the longitudinal and transverse part\cite{ks1}.
Taking $\vct{q}=(q,0,0)$,
the former is the $3\times 3$ matrix depending on $a=-1,0$ and $1$,  
while the latter is composed of $a=2$ and $3$.
The longitudinal part, which is required for the present discussions,
is calculated in NFA and NSA as
\begin{equation}
 D_{\mathrm{L}}=
\mathrm{det}\Bigl(\delta_{ab}-\kappa_a R_{ab}(\omega)\Bigr)
=
\left|
\begin{array}{ccc}
  C_\sigma+f_\sigma \rph_\sigma
& f_\sigma \rph_{\sigma\omega} 
& sf_\sigma \rph_{\sigma\omega} \\[4pt]
 -\,f_\omega \rph_{\sigma\omega}
& C^0_\omega-f_\omega \rph_{\omega}
& -\,sf_\omega \rph_{\omega} \\[4pt]
 sf_\omega \rph_{\sigma\omega}
& sf_\omega \rph_{\omega}
& C^1_\omega+f_\omega  s^2\rph_{\omega}
\end{array}
\right|\label{dl}
\end{equation}
where we have used the following abbreviations,
\begin{equation}
C_\sigma=1-f_\sigma\rhn_{-1-1}\,,\qquad
C^0_\omega=1+f_\omega\rhn_{0 0}\,,\qquad
C^1_\omega=1-f_\omega\left( \rhn_{1 1}+\frac{\nf\vf^2}{3} \right),
\label{diagonal}
\end{equation}
together with $s=\omega/q$, $\vf = \pf/\epf$
for the  relativistic Fermi velocity 
and $\nf=2\vf \epf^2/\pi^2$
for the relativistic density of states at the Fermi surface.
Here $\epf$ is defined by $\sqrt{\pf^2+M^{\ast 2}}$, 
using the Fermi momentum $ \pf$.
The function $\rph$ for isosymmetric nuclear matter is given by
\[
  \rph_{\sigma} 
=\nf\left( 1-\vf^2 \right)\varPhi(x) \\[3pt], \quad
\rph_{\sigma\omega}
=\nf \sqrt{1-\vf^2}\,\varPhi(x) \\[2pt], \quad
\rph_{\omega} =\nf\varPhi(x),
\]
where $\varPhi(x)$ stands for Lindhard function with $x=\omega/\vf q$,
\begin{equation}
\varPhi(x)=
-\,1+\frac{x}{2}\log\left|\frac{1+x}{1-x}\right|
-\frac{i\pi}{2}|x|\,\theta(\,1-|x|\,).
\label{def_eq_phi}
\end{equation}
By writing the determinant Eq.(\ref{dl}) in terms of the Landau-Migdal
parameters, $F_0$ and $F_1$, as \cite{ks6}
\begin{align}
 \frac{D_{\mathrm{L}}}{C_\sigma C^0_\omega C^1_\omega}
&=
 1+\frac{f_\sigma}{C_\sigma}\rph_\sigma
-\frac{f_\omega}{C^0_\omega}\rph_\omega
+\frac{f_\omega}{C^1_\omega}s^2\rph_\omega \nonumber\\
&= 
1-\left(F_0+\frac{F_1}{1+F_1/3}x^2\right)\varPhi(x)\, ,
\end{align}
we obtain 
\begin{equation}
F_0=-  \frac{(1-\vf^2)F_\sigma}{1-f_\sigma\rhn_{-1-1}}
 +\frac{F_\omega}{1+f_\omega\rhn_{00}}\ ,\quad
 F_1=-\frac{\vf^2F_\omega}{1-f_\omega\rhn_{11}}\ ,\\[4pt]
\end{equation}
with $F_\omega = \nf f_\omega$ and $F_\sigma = \nf f_\sigma$.
Thus, Pauli blocking terms yield the denominators depending on
each meson exchange. They are calculated as,
\begin{equation}
 \rhn_{-1-1}=-
4 \int_0^{\pf}\!\!\frac{\dd^3p}{(2\pi)^3}
\frac{\vct{p}^2}{E^3_{\vct{p}}} , \qquad
\rhn_{00}=0\,, \qquad
\rhn_{11}=-
4\int_0^{\pf}\!\!\frac{\dd^3p}{(2\pi)^3}
\frac{E^2_{\vct{p}}-\vct{p}^2/3}{E^3_{\vct{p}}},\label{rhn}
\end{equation}
It is seen that the attractive interactions are quenched
by the Pauli blocking terms 
in the same way as in Eq.(\ref{NSA_kappa}).
In the present model, there is no contribution to the repulsive part,
since $\rhn_{00}=0$ due to $f_{\alpha\neq \beta}^{(0)}=0$.

The parameter $F_0$ is responsible for the
nuclear incompressibility coefficient, which determines 
the restoring force of giant monopole state. Reduction of the attractive
part makes the value of the coefficient higher \cite{ks2}.

In contrast to $F_0$, the parameter $F_1$ is constrained
by more fundamental requirement that the Femi energy $\ef$
and momentum $\pf$ are transformed as a four-vector \cite{baym}. 
In the $\sigma$-$\omega$ model, 
the parameter $F_1$ comes from the longitudinal part
of the $\omega$-meson exchanges as a relativistic effect,
while the nucleon effective mass stems from the $\sigma$-meson exchange.
They, however, are not independent of each other, as in nonrelativistic
models, and should satisfy, according to the above requirement\cite{ks4},
\begin{equation}
\frac{\epf}{\ef}=1+\frac{F_1}{3}.\label{f1}
\end{equation}
for $\ef=E_{\mathrm{B}}+M$, $E_{\mathrm{B}}$
being the binding energy per nucleon.
As far as Eq.(\ref{f1}) holds,  $F_1$ describes correctly
the center of mass motion by the Lorentz boost\cite{nks}, 
and restores also an abnormal enhancement of magnetic moments
due to the effective mass in the Hartree approximation\cite{ks4}.
Thus, although NFA and NSA seem to be consistent with the framework 
of the Landau-Migdal parameters, this fact does not imply that
the divergence can be neglected.

We note that the last term $\nf\vf^2/3$ in $C^1_\omega$
of Eq.(\ref{diagonal}) comes from
particle-hole excitations through the space component
of the $\omega$-meson exchange
\begin{equation}
\rph_{11} =s^2\rph_\omega-\frac{\nf\vf^2}{3} .
\label{rph_omega_11}
\end{equation}
Because of the last term, the continuity equation 
does not hold in the particle-hole space,  
$\rph_{11} \neq s^2\rph_{00}=s^2\rph_\omega$.
NFA and NSA, however, take into account a part of N-$\nbar$ states.
The calculation of $\rhn_{11}$ in Eq.(\ref{rhn}) provides 
$-\nf\vf^2/3$, which leads to the continuity
equation, $\rph_{11}-\rhn_{11}=s^2(\rph_{00}-\rhn_{00})$,
as discussed in \S\ref{ce}.

\section{Conclusions}

The structure of the relativistic random phase approximation(RPA) has 
been investigated in detail.
The energy-weighted sum of the RPA transition strengths
is expressed formally as the Hartree ground-state
expectation value of the double commutator between the excitation
operator and the Hamiltonian, as in non-relativistic models.
In calculating the commutator, however, the usual anticommutation relation
between the baryon fields cannot be used\cite{ks5}.
Otherwise, the sum, which should be infinite\cite{ks5}, would vanish.

The main difference of the relativistic RPA from the nonrelativistic one
stems from antinucleon($\nbar$) degrees of freedom,
but they cause the divergence problems. 
The two kinds of approximations were proposed by previous authors
in order to avoid the problems without the renormalization.
The one\cite{chin,ks1} is
the no free term approximation(NFA) which simply neglects the
divergent terms in RPA response function,
and the other\cite{df} is the no-sea approximation(NSA)
where $\nbar$ states are assumed to be empty. 
Actually, both approximations are equivalent to each other.
They were employed widely and shown to work well for reproducing
experimental data in a phenomenological way\cite{lnnm,ks1,srm,df,piek,mgwvr,mcg}.

The present paper has shown that NFA and NSA 
have the serious problems.
The RPA dispersion equation yields the RPA states
with negative excitation energy,
in addition to the low lying positive energy states.
This fact implies
that the RPA ground state is not the lowest one. 
Owing to those negative excitation-energy states,
the energy-weighted sum of the transition strengths vanishes.
These results are not avoidable for NFA and NSA which satisfy
the RPA relation of the energy-weighted strengths,
since the relativistic sum rule value stems from
the excitations of Dirac sea\cite{ks5}.  
Moreover, since the only limited space of nucleon-antinucleon states
is included in NFA and NSA, attractive(repulsive) forces work as
repulsive(attractive) ones between their couplings.
This fact affects also the couplings of the particle-hole states
with nucleon-$\nbar$ states.
Unfortunately, these unphysical couplings played an important role
in explaining the spurious state\cite{df} 
and the giant monopole states\cite{piek,mgwvr,mcg}
in the previous numerical calculations in NSA.
These results have been shown clearly by using a schematic model.

It has been shown that there is no problem
for NSA and NFA to describe the continuity equation,
since it is independent of the divergence. 

Thus, $\nbar$ degrees of freedom which provide the divergence
are not ignored. As far as a part of the $\nbar$ space
is necessary, the rest of the space also should be taken into account
in a proper way, even in phenomenological models.
Indeed, it was shown in Refs.\cite{chin,ks2,ks7,ks8}
that the renormalization of the divergence plays an important role
in discussions of some physical quantities.
Those roles are state-dependent, 
and could not be incorporated into phenomenological 
interactions or their coupling constants.
Moreover, if the divergence of the linearly energy-weighted sum
is understood, we can make clear the meaning of the analyses as to 
the distribution of transition strengths with the energy moments\cite{cpc}.

\end{document}